\documentclass[twocolumn,preprintnumbers,amsmath,amssymb]{revtex4}
\usepackage{graphicx}
\usepackage{dcolumn}
\usepackage{bm}
\usepackage[dvips]{color}

\newcommand{\beq}{\begin{equation}}
\newcommand{\eeq}{\end{equation}}
\newcommand{\bq}{\begin{equation}}
\newcommand{\eq}{\end{equation}}
\newcommand{\ba}{\begin{array}}
\newcommand{\ea}{\end{array}}
\newcommand{\beqa}{\begin{eqnarray}}
\newcommand{\eeqa}{\end{eqnarray}}

\def\f{\frac}
\def\dis{\displaystyle}
\def\[{\left[}
\def\]{\right]}
\def\({\left(}
\def\){\right)}
\def\ga{g_0^{~}}
\def\gaa{g_0^2}
\def\gb{g_1^{~}}
\def\gbb{g_1^2}
\def\gc{g_2^{~}}
\def\gcc{g_2^2}
\def\fa{f_1^{}}
\def\faa{f_1^2}
\def\fb{f_2^{}}
\def\fbb{f_2^2}

\def\pslash{\not{\hbox{\kern-4pt $p$}}}
\def\qslash{\not{\hbox{\kern-4pt $q$}}}
\def\lv{\not{\hbox{\kern-4pt $L$}}}
\def\lsim{\mathrel{\raise.3ex\hbox{$<$\kern-.75em\lower1ex\hbox{$\sim$}}}}
\def\gsim{\mathrel{\raise.3ex\hbox{$>$\kern-.75em\lower1ex\hbox{$\sim$}}}}
\def\ifmath#1{\relax\ifmmode #1\else $#1$\fi}
 \evensidemargin=\oddsidemargin

\begin{document}

 \title{\Large LHC Signatures of New Gauge Bosons
               in the Minimal Higgsless Model}

 \author{Hong-Jian He$^{1}$, Yu-Ping Kuang$^{1}$,
         Yong-Hui Qi$^{1}$, Bin Zhang$^{1}$}

 \affiliation{
 $^{1}$Center for High Energy Physics, Tsinghua
       University, Beijing 100084, China }

 \author{Alexander Belyaev$^{2}$, R.\ Sekhar Chivukula$^{2}$,
         Neil D.\ Christensen$^{2}$,
         Alexander Pukhov$^{3}$, Elizabeth H.\ Simmons$^{2}$}

 \affiliation{
 $^{2}$Department of Physics and Astronomy, Michigan State
       University, East Lansing, MI 48824, USA\\
 $^{3}$Skobltsyn Institute of Nuclear Physics,
       Moscow State University, Moscow 119992, Russia}

 \begin{abstract}
 We study the LHC signatures of new gauge bosons in the gauge-invariant
 minimal Higgsless model. It predicts an extra pair of $W_1$ and $Z_1$
 bosons which can be as light as $\sim\!400$\,GeV and play a key role in
 the delay of unitarity violation. We analyze the $W_1$ signals in
 $pp\to W_0Z_0Z_0\to jj4\ell$ and
 $pp\to W_0Z_0jj\to \nu 3\ell jj$ processes at the LHC, including the
 complete electroweak and QCD backgrounds. We reveal the complementarity
 between these two channels for discovering the $W_1$ boson,
 and demonstrate the LHC discovery potential over the full
 range of allowed $W_1$ mass.
 \\[2mm]
  PACS: 12.60.Cn, 11.10.Kk, 12.15.Ji, 13.85.Qk  \hfill
  {arXiv:\,0708.2588}~[hep-ph]
 \end{abstract}

 \maketitle

 \noindent
 {\bf 1.~Introduction}
 \\[3mm]
 Unraveling the mechanism of electroweak symmetry breaking (EWSB) is the
 most pressing task facing particle physics today, and is a major driving
 force behind the CERN Large Hadron Collider (LHC). New spin-1 gauge bosons
 serve as the key for Higgsless EWSB\,\cite{terning1}, by delaying unitarity
 violation of longitudinal weak boson scattering up to a higher ultraviolet
 scale \cite{unitary5D,HeDPF04} without invoking a fundamental Higgs scalar
 \cite{higgs}. Dimensional deconstruction \cite{DC} was shown to provide
 the most general {\it gauge-invariant} formulation \cite{HeDPF04,DC04a} of
 Higgsless theories under arbitrary geometry of the continuum fifth dimension
 (5d) or its 4d discretization with only a few lattice sites
 \cite{4site,3site}. The Minimal Higgsless Model (MHLM) consists of just 3
 lattice sites (``The Three Site Model'')\,\cite{3site},
 and includes extra nearly degenerate
 $(W_1,\,Z_1)$ bosons which are allowed by precision data
 to be as light as $\sim\!400$\,GeV \cite{3site}.
 The model contains all the essential ingredients of Higgsless theories,
 and is the simplest realistic Higgsless model with distinct collider
 signatures.
 In this paper we study the LHC signatures of the new $W_1$ boson
 via processes
 $\,pp\to W_0Z_0Z_0\to jj4\ell\,$ and
 $\,pp\to W_0Z_0 jj\to \nu 3\ell jj\,$,
 where $(W_0,\,Z_0)$ are the light weak gauge bosons analogous to
 those in the standard model (SM).
 The MHLM is exactly gauge-invariant
 with spontaneous gauge symmetry breaking,
 hence it allows our analysis to predict consistent high-energy
 behavior for any relevant scattering amplitude,
 contrary to the previous gauge-noninvariant calculation in a
 naive 5d Higgsless sum rule approach\,\cite{mat}.

 \vspace*{5mm}
 \noindent
 {\bf 2.~Model Setup and Delayed Unitarity Violation\,}
 \\[3mm]
 The MHLM\,\cite{3site} is defined as a
 chain moose with 3 lattice sites, under the gauge groups
 \,$SU(2)_0\otimes SU(2)_1\otimes U(1)_2$ \cite{Casalbuoni}.\,
 Its gauge and Goldstone sectors have 5 parameters in total ---
 3 gauge couplings $(\ga,\,\gb,\,\gc)$ and 2 Goldstone decay constants
 $(\fa,\,\fb)$, satisfying two conditions due to
 its symmetry breaking structure,
 \beqa
 \hspace*{-6mm}                          
 \f{1}{\gaa} + \f{1}{\gbb} + \f{1}{\gcc} ~=~ \f{1}{e^2}\,,
 ~&&~
 \f{1}{\faa} + \f{1}{\fbb} ~=~ \f{1}{v^2}\,.
 \eeqa
 For the optimal delay of unitarity violation
 we will choose equal decay constants
 $\,\fa =\fb=\sqrt{2}v$\, where $\,v=\(\sqrt{2}G_F\)^{-1/2}\,$
 as fixed by the Fermi constant.
 Inputting two gauge boson masses, e.g.,
 the light gauge boson mass $M_{W0}\simeq 80$\,GeV
 and new gauge boson mass
 $M_{W1}$, we can determine all three gauge couplings
 $(\ga,\,\gb,\,\gc)$.

 The MHLM exhibits a delay of unitarity violation
 in light weak boson scattering
 $V_{0L}^aV_{0L}^b\to V_{0L}^cV_{0L}^d$ ($V_0=W_0,Z_0$).\,
 The scattering $W_{0L}Z_{0L}\to W_{0L}Z_{0L}$
 is related to our collider study, so we
 derive the corresponding unitarity limits,
 $E^\ast \simeq 3.1,\,2.95,\,2.8,\,2.55,\,2.35,\,1.7$\,TeV for
 $M_{W1}=0.4,\,0.5,\,0.6,\,0.8,\,1.0,\,\infty$\,TeV, respectively.
 (We have also analyzed the unitarity limit
 in the combined isospin-0 channel
 and find somewhat tighter limits:
 $E^\ast = 2.0,\,1.86,\,1.74,\,1.66,\,1.45,\,1.2$\,TeV, for
 $M_{W1}$$=0.4,\,0.5,\,0.6,\,0.8,\,1.0,\,\infty$\,TeV, respectively.)
 So, for $M_{W_1}\lesssim 1$\,TeV,
 each elastic $V_{0L}V_{0L}$ scattering remains
 unitary over the main energy range of the LHC.

 The fermion sector contains SM-like chiral fermions: left-handed
 doublets $\psi_{0L}^{}$ under $SU(2)_0$ and right-handed weak
 singlets $\psi_{2R}^{}$. For each flavor of $\psi_{0L}^{}$, there is
 a heavy vector-fermion doublet $\Psi_1^{}$ under $SU(2)_1$.
 The mass matrix for $\{\psi,\Psi\}$ is\,\cite{3site}
 \beq                                   
 \dis
 M_F = \left(\ba{cc} m & 0 \\[2mm]
                     M & m' \ea\right)
 \equiv M\left(\ba{cc} \epsilon_L^{} & 0 \\[2mm]
                                   1 & \epsilon_R^{}  \ea\right) \,.
 \eeq
 The light SM fermions acquire small masses proportional to
 $\epsilon_R^{}$\,.\,
 For the present high energy scattering analysis
 we only need to consider light SM fermions
 relevant to the proton structure functions at the LHC, which
 can be treated as massless to good accuracy. So we will set
 \,$\epsilon_R^{}\simeq 0$,\, implying that
 $\psi_{2R}$ and $\Psi_{1R}$ do not mix.
 The mass-diagonalization of $M_{F}$ yields
 a nearly massless SM-like light fermion $F_0$
 and a heavy new fermion $F_1$ of mass
 $\,M_{F_1} = M\sqrt{1+\epsilon_L^2}\,$.
 The low energy constraints already put a strong lower bound on
 the heavy fermion mass, $\,M_{F_1} > 1.8$\,TeV \cite{3site}.
 Hence we focus our analysis on the production and detection
 of the new gauge boson $W_1\,(Z_1)$
 which can be as light as $\sim\!400$\,GeV.
 To simplify the analysis we will consistently
 decouple the heavy fermion by taking the limit
 $\,(M,\,m)\to\infty\,$ while keeping the ratio
 $\,\epsilon_L^{}\equiv m/M$ finite.
 This finite ratio $\epsilon_L^{}$ will be fixed via the ideal
 fermion delocalization\,\cite{ideal},
 leading to vanishing $W_1$-fermion couplings and
 thus zero electroweak precision corrections
 at tree-level\,\cite{3site,ideal}.
 We stress that the precisely defined fermion gauge couplings
 as well as gauge-boson self-couplings
 in the MHLM\,\cite{3site} are the key to ensuring exact
 gauge-invariance in our collider study below.
 This feature makes our analysis essentially different from
 Ref.\,\cite{mat} which relies on a naive 5d Higgsless
 sum rule approach where both the fermion gauge couplings and
 the deviations of the $W_0/Z_0$ self-couplings from the SM
 cannot be correctly derived.

 \vspace*{5mm}
 \noindent
 {\bf 3.~Discovering the $\boldsymbol{W_1}$ Boson via
 $\boldsymbol{pp \to W_0Z_0Z_0}$}
 \\[3mm]
 Since $W_1$ has vanishing fermionic couplings in the
 MHLM due to ideal fermion-delocalization\,\cite{3site,ideal},
 it decays to $W_0Z_0$ only, with the total width,
 \beqa                                      
 \label{eq:W1-width1}
 \Gamma_{W_1}
 &=& \f{\,\alpha M_{W_1}^3\,}
   {192s_Z^2M_{W_0}^2}\!\!
   \[1\!+\!\f{\;9\!+\!7c_Z^2\;}{c_Z^2}r^2 +O(r^4)\]\,
 \eeqa
 where $\,\alpha =e^2/4\pi$\, and \,$r\equiv M_{W0}/M_{W1}\ll 1$.\,
 We note that in Eq.\,(\ref{eq:W1-width1})
 the total width at the leading order $(r^0)$ comes from
 the purely longitudinal decay mode $W_{1L}\to W_{0L}Z_{0L}$\, alone.
 Typically, we have $\,\Gamma_{W_1}\simeq (3,\, 5,\, 17,\, 31)\,$\,GeV
 for $M_{W_1}=(0.4,\,0.5,\,0.8,\,1)$\,TeV. 

 In this and next sections our collider studies will be performed at the
 parton-level (unless specified otherwise).
 Now we study the new process $pp \to W_0Z_0Z_0$ at the LHC,
 where the signal comes from
 \,$pp \to W_0^\ast\to W_1^{(\ast)}Z_0\to W_0Z_0Z_0$\,.\,
 We propose to analyze the final state detection via
 leptonic decays of the two $Z_0$ bosons and
 hadronic decays of $W_0$. This gives rise to a clean signature
 of 4-leptons plus 2-jets,
 \,$\,jj\ell^+\ell^-\ell^+\ell^-$\, ($\ell=e,\mu,~jj=qq'$).
 The backgrounds include:
 (a) the irreducible SM production of
 $pp\to W_0Z_0Z_0\to jj\ell^+\ell^-\ell^+\ell^-$
 ($jj=qq'$, without $W_1$ contribution),
 (b) the reducible background of the SM production,
 $\,pp\to Z_0Z_0Z_0\to jj\ell^+\ell^-\ell^+\ell^-\,$ ($jj=q\bar q$),
 with one $Z_0\to jj$ (mis-identified as $W_0$)
 since the mass-splitting $M_{Z_0}-M_{W_0}$ is
 within the uncertainty of reconstructing the $W_0$ boson,
 and (c) the SM process $pp\to jj\ell^+\ell^-\ell^+\ell^-$
 other than (a) and (b),
which
 also includes the $jj 4\ell$ backgrounds with $jj=qg,gg$.\,

 We first impose the following cuts for particle identifications
 \begin{eqnarray}                          
 p_{T\ell}^{~}>10\,{\rm GeV},~~&&~~ |\eta_\ell^{~}| < 2.5\,,
 \nonumber\\
 p_{Tj}^{~}>15\,{\rm GeV},~~&&~~ |\eta_j^{~}| < 4.5\,.
\label{det-cuts}
 \end{eqnarray}
 To suppress the backgrounds, we further impose
 \begin{eqnarray}                            
 && M_{jj} \,=\, 80\pm 15\,{\rm GeV},
 ~~~~~~\Delta R(jj)\,<\,1.5\,,
 \nonumber\\
 && \sum_{i=1}^2 p_T^{~}(Z_i)+\sum_{i=1}^2 p_T^{~}(j_i)
 \,=\, \pm 15~{\rm GeV}.
 \label{supp-cuts}
 \end{eqnarray}
 The $M_{jj}$ cut is the requirement of reconstructing $W_0$ from the dijets
 (due to on-shell $W_0$ decay)
 according to the experimental resolution $\pm 15$ GeV\,\cite{ATLAS}.
 The $\Delta R(jj)$ cut is for suppressing the \,$jj=qg,gg$\, backgrounds.

\begin{figure}[h]
 \vspace*{-12mm}
 \hspace*{-5mm}
 \includegraphics[width=9.4cm,
 ]{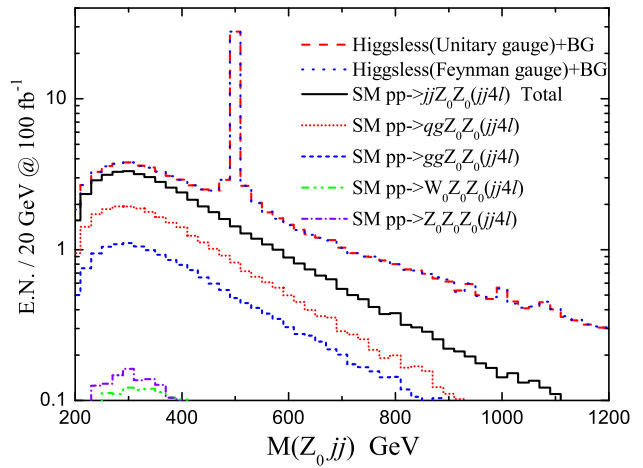}
 \vspace*{-12mm}
 \caption{Signal and background events in the process
 $pp\to W_1^{(\ast)}Z_0 \to W_0Z_0Z_0\to jj\,\ell^+\ell^-\ell^+\ell^-$
 for an integrated luminosity  of 100\,fb$^{-1}$.}
  \label{M(WZ)2}
 \vspace*{-2mm}
 \end{figure}

 In practical data analyses for reconstructing $W_0$,
 there is no need to require separation between the two jets.
 Then there can be a background from a 
 single jet or overlapping jet pair (arising from a single parton) with invariant-mass satisfying the $M_{jj}$ requirement
 (\ref{supp-cuts})  which may mimic the two unseparated jets.
 So we need to impose a further cut to suppress this background.
 With a detailed simulation study using PYTHIA with the ${\rm k_T}$ algorithm (taking the default
 parameter $d_{cut}=\sqrt{\hat s}$) and considering showering effects,
 we find that the cut requiring $p_T^{~}$ balance
 can suppress the single-jet background to the level of $10^{-2}$ events which is invisible in
 FIG.\,\ref{M(WZ)2}.

 The relevant SM backgrounds are simulated using the Madgraph package\,\cite{MADGRAPH}
 and a few related ones\,\cite{CalcHEP}.

\begin{figure}[b]
 \vspace*{-11mm}
 \hspace*{-6mm}
 \includegraphics[width=9cm, height=7cm
  ]{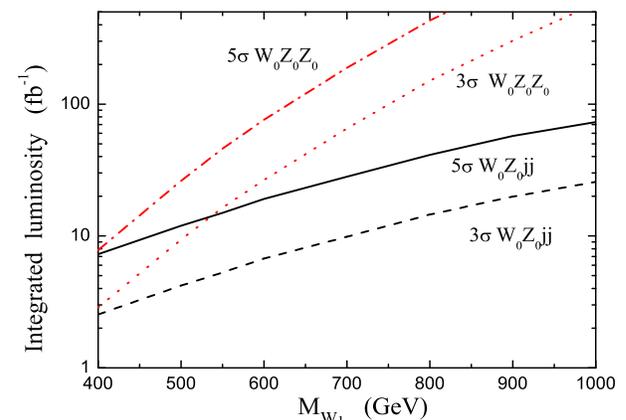}
 \vspace*{-9mm}
 \caption{Integrated luminosities required
 for $3\sigma$ and $5\sigma$ detection
 of $W_1$ signals as a function of $M_{W_1}$.
 The dotted and dashed-dotted curves are for the $W_0Z_0Z_0$
 channel, while the dashed and solid curves are
 for the $W_0Z_0jj$ channel.}
 \label{IL}
 \vspace*{2mm}
 \end{figure}

 We plot the signal and background events
 under these cuts for an integrated luminosity of 100\,fb$^{-1}$
 in Fig.\,\ref{M(WZ)2}, where we depict the signal by a dashed curve,
 the backgrounds (c) with $jj=gg,qg$ by small-dashed and
 small-dotted curves, respectively.
 The backgrounds (a) and (b) with $jj=qq,q\bar{q}$ are much smaller, appearing
 marginally on the left-lower corner of Fig.\,\ref{M(WZ)2}.
 We summarize the total backgrounds (a)$+$(b)$+$(c),
 depicted by the solid curve in Fig.\,\ref{M(WZ)2}.
 We define the signal to include all events
 in the region \,$M_{Zjj}=M_{W_1}\pm 0.04M_{W_1}$,\,
 where the backgrounds are much smaller than the signal.
 The final state $W_0Z_0Z_0$
 contains two identical $Z_0$ bosons and
 we have summed up the contributions from the
 two combinations of $M_{Zjj}$ for all the curves in
 Fig.\,\ref{M(WZ)2}.
 For the signal curve, the non-resonant region on the
 right-hand-side of the peak mainly comes from the
 contributions of the two $Zjj$-combinations including
 no $W_1$ peak; this region also decreases more slowly
 than the other backgrounds.
 The gauge-invariance of this calculation is verified by
 comparing the signal distributions in unitary and
 't\,Hooft-Feynman gauges; as shown in Fig.\,\ref{M(WZ)2}
 by red-dashed and blue-dotted curves, they perfectly
 coincide. We further derive statistical significance
 from Poisson probability in the conventional way.
 The integrated luminosity required for detecting the new $W_1$ gauge bosons
 in this channel will be summarized in Fig.\,\ref{IL}
 as a function of mass $M_{W_1}$.


 \vspace*{3mm}
 \noindent
 {\bf 4.~Discovering the $\boldsymbol{W_1}$ Boson via
 $\boldsymbol{pp \to W_0Z_0jj}$}
 \\[3mm]
 Next, we analyze the discovery of $W_1$ bosons in the
 process $pp \to W_0Z_0qq'$, where the signal is
 given by the $W_1$ contribution to the
 $W_0Z_0\to W_0Z_0$ subprocess.
 We perform a complete analysis of
 \,$pp\to W_0Z_0jj$, and choose the pure leptonic decay
 modes of $W_0Z_0$ with 3 leptons plus missing-$E_T$
 \cite{bagger,wwh}.
 We carry out the first full $2\to 4$ calculation for the MHLM,
 including both the electroweak (EW) and QCD backgrounds for
 \,$pp\to W_0Z_0jj$.
 We find the total SM backgrounds to be an order of magnitude
 larger than those estimated in \cite{mat}.

 We first analyze how to suppress the SM backgrounds.
 We employ a forward-jet tag to eliminate
 the annihilation process $qq' \to W_0Z_0$ (with possible
 QCD-jet radiation), as a
 reducible QCD background\,\cite{bagger}.
 Of greater concern are the reducible
 QCD backgrounds
 $pp \to W_0Z_0 jj$ with $jj = qg,\,gg$ serving as forward
 jets. We find that these two QCD backgrounds are quite
 significant even under the cuts of \cite{bagger,mat}.
 Hence we employ the following improved cuts to more
 effectively suppress the backgrounds,
 \begin{eqnarray}                           
 \vspace*{-5.5mm}
 \hspace*{-10mm}
 E_j^{}    > 300\,{\rm GeV}\,,~~&&~~
 p_{Tj}^{} >  30\,{\rm GeV}\,,
 \label{oldcuts}
 \\
 |\eta_j^{}| < 4.5\,,
 \hspace*{7mm}
 ~~&&~~
 \left|\Delta\eta_{jj}^{}\right| > 4\,,
 \label{jj rap diff}
 \label{newcuts}
 \end{eqnarray}
 where $E_j^{}$ and $p_{Tj}^{}$ are transverse energy and
 momentum of each final-state jet,
 $\eta_{j}^{}$ is the forward jet rapidity, and
 $\left|\Delta\eta_{jj}^{}\right|$ is
 the difference between the rapidities of the two forward
 jets. The cut on $|\Delta \eta_{jj}^{~}|$
 suppresses\,\cite{HLLHCpaper2}
 the QCD backgrounds \,$pp \to W_0Z_0gg,~W_0Z_0qg$,\,
 especially in the low $M_{W_0Z_0}$ region.
 In addition, we employ the following lepton
 identification cuts,
 \null\vspace{-0.5cm}
  \begin{equation}                        
 p_{T\ell}^{~} \,>\, 10\,{\rm GeV}\,, ~~~~~~~
 |\eta_{\ell}^{~}| \,<\, 2.5\,.
 \label{newcuts-lepton}
 \vspace*{-4mm}
 \end{equation}
 We have also included the irreducible QCD backgrounds to $pp\to W_0Z_0qq'$
 (cf.\ dashed curve in Fig.\,3 below).

 For computing the SM EW backgrounds, we need to
 specify the reference value of the SM Higgs
 mass $M_H$. Because the SM Higgs scalar only contributes
 to the $t$-channel in $pp\to qq'W_0Z_0 $, we find that
 varying the Higgs mass in its full range
 $M_H=115\,{\rm GeV}-1\,$TeV has little effect on the SM
 background curve.
 Hence we can simply set $M_H=115$\,GeV in our
 plots without losing generality.

  The process $pp \to W_0 Z_0 qq'$ is a $2 \to 4$ scattering
 process including both fusion and non-fusion contributions,
 where the former are the diagrams with a fusion sub-process
 $W_0 Z_0 \to W_0 Z_0$.
 Using the improved cuts (\ref{oldcuts})-(\ref{newcuts-lepton}),
 we have computed the $W_0 Z_0$ invariant mass ($M_{W_0Z_0}$)
 distribution in both unitary gauge and
 't\,Hooft-Feynman gauge as depicted in Fig.\,2(a).
 The final result (the sum of the fusion and non-fusion
 pieces) is identical in the two gauges, as shown in
 Fig.\,2(b).

 \begin{widetext}
 \begin{center}
 \begin{figure}[h]
 \vspace*{-3mm}
 \hspace*{-2mm}
 \includegraphics[width=18cm,height=5.5cm]{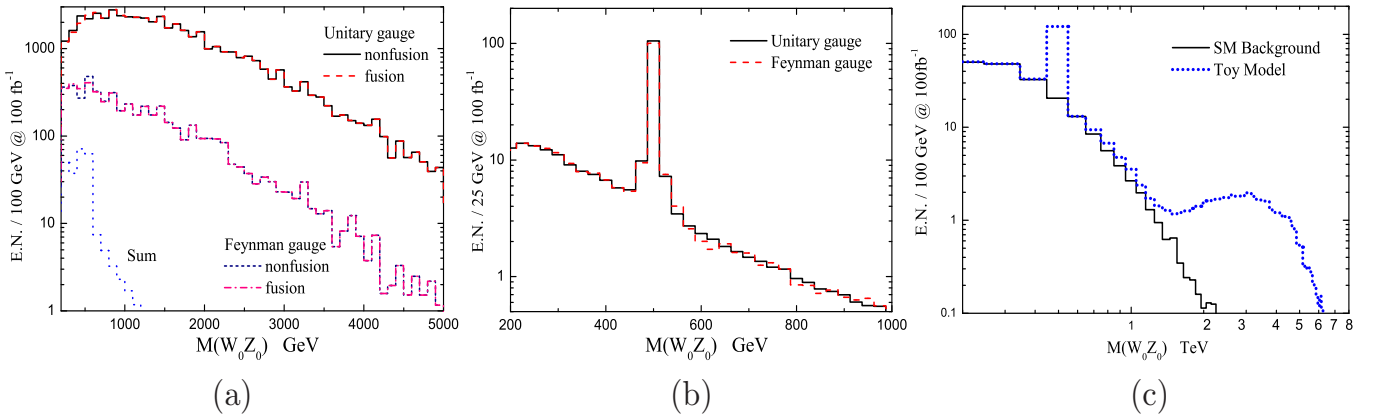}
 \vspace*{-6mm}
 \caption{Invariant-mass distribution
 $M_{W_0Z_0}$ in $pp\to W_0Z_0qq'$ for
 $M_{W_1}=500$\,GeV: (a).\ fusion (dashed) and non-fusion (solid)
 contributions in the unitary gauge,
 fusion (densed-dashed) and non-fusion (dashed-dotted) contributions
 in the 't\,Hooft-Feynman gauge,
 and the sum of fusion and non-fusion contributions in
 both gauges which coincide (dotted);
 (b).\ comparison of the summed contributions (after cancellation)
 between the two gauges, in the MHLM;
 (c).\ same as (b), but for a toy model based on a sum-rule approach (like that in Ref.~\cite{mat})
 which explicitly violates gauge-invariance.}
 \vspace*{-4mm}
 \label{fig:Cancel-UF}
 \end{figure}
 \end{center}
 \end{widetext}

 Strikingly, we have revealed an extremely precise and large
 cancellation between the fusion and non-fusion contributions,
 as required by the exact gauge-invariance of
 {\it the MHLM Lagrangian}\,\footnote{It is well-known
 that in any perturbative expansion up to a finite order,
 the width of an $s$-channel resonance
 (such as $W_1$ in the MHLM or Higgs boson in the SM) will violate gauge-invariance
 due to neglecting the higher order effect.
 But our study explicitly shows that, for the small $\Gamma_{W_1}$ in
 (\ref{eq:W1-width1}) as predicted by the MHLM,
 such an effect is not visible up to energies well above 1\,TeV
 in current numerical analysis.}.
 This manifest gauge-invariance of the MHLM Lagrangian is crucial for obtaining
 the correct collider phenomenology and nontrivially verifies the consistency of our
 analysis. Note that the separate fusion and non-fusion
 contributions are gauge-dependent and a large precise cancellation occurs
 {\it only if they are rigorously combined,} i.e.,
 all the new physics contributions to the gauge-couplings of $W_0/Z_0$
 and $W_1/Z_1$ as well as their couplings to the light fermions
 {\it must be consistently included}. The curves shown in Fig.\,2(a)(b)
 come from the Higgsless model and no SM Higgs boson is invoked\,\footnote{
 This contrasts with the case of the SM, where Higgs-exchange is crucial to cancel the
 order $E^2$ fusion-diagram contributions to $WZ \to WZ$.}.
 To be concrete, we see that at $M_{W_0Z_0}=1\,$TeV, the cancellation in
 the unitary gauge is a factor of \,$2400/2.3 \simeq 1043$,\, while that
 in 't\,Hooft-Feynman gauge is a factor of
 \,$195/2.3 \simeq 84.7$.\,
 We stress that these cancellations cannot be inferred
 without a truly gauge-invariant model.

 As a comparison, we also show, in Fig.\,2(c), the summed result of the fusion and
 non-fusion contributions in unitary gauge for a naive toy model that is not gauge
 invariant. Following the 5d sum rule approach\,\cite{mat}, we assume that
 $W_0$ and $Z_0$ have exactly the SM-couplings to
 light fermions while $W_1$ and $Z_1$ do not couple to the fermions. Next, one has to
 estimate the gauge boson self-couplings. There are at least 3 self-gauge-couplings (involving
 $W_0$-$W_0$-$Z_0$-$Z_0$, $W_0$-$W_0$-$Z_0$ and $W_1$-$W_0$-$Z_0$ vertices)
 which cannot be equal even after assuming all higher KK modes are fully decoupled. But only two sum rules
 from requiring $E^4$ and $E^2$ cancellations in $W_0Z_0\to W_0Z_0$ can be derived;
 the $E^2$ sum rule may be used to estimate the $W_1$-$W_0$-$Z_0$ coupling (ignoring all higher KK modes)
 and then the $E^4$ sum rule could determine the $W_0$-$W_0$-$Z_0$ coupling if one assumes the
 $W_0$-$W_0$-$Z_0$-$Z_0$ coupling equals the SM-value.  With this setup, we recompute
 Fig.\,2(a) and plot the summed results in Fig.\,2(c); it shows that the precise cancellation we
 found for the MHLM in Fig.\,2(b) is now destroyed
 in the high energy region $1.5-7$\,TeV, causing a large fake peak around 3\,TeV in Fig.\,2(c).

 \begin{center}
 \begin{figure}[h]
 \vspace*{-11.5mm}
 \hspace*{-5mm}
 \includegraphics[width=9cm, height=7cm
 ]{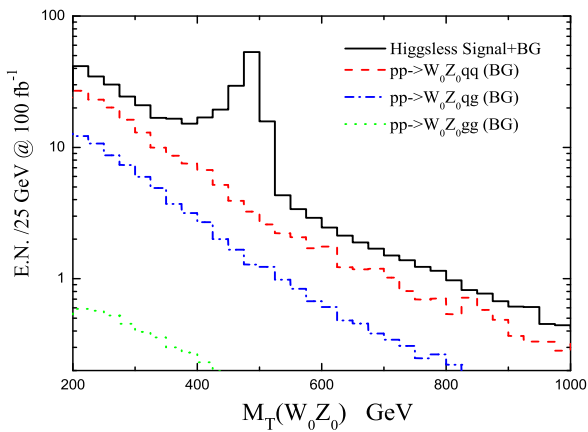}
 \vspace*{-11mm}
 \caption{Numbers of signal and background events versus
 the transverse mass $M_T(W_0Z_0)$ after imposing the cuts
 (\ref{oldcuts})-(\ref{newcuts-lepton})
 for an integrated luminosity of 100\,fb$^{-1}$.}
  \vspace*{-5mm}
 \label{sb-newcuts}
 \end{figure}
 \end{center}

 \null\vspace{-0.5cm}Traditional analyses\,\cite{bagger} of gauge-boson fusion
 in a strongly-interacting symmetry breaking sector have
 relied on using separate calculations of the signal and
 background. The background is calculated in the SM, while
 the signal could be calculated only by using a model of
 $2\to 2$ Goldstone-boson scattering and
 applying the equivalence theorem together with
 effective-$W$ approximation.
 The MHLM Lagrangian\,\cite{3site} is exactly gauge-invariant
 and, as we have demonstrated here, allows direct
 calculations of full $2\to 4$ processes in any gauge.

 \vspace*{2mm}
 In a realistic experimental analysis,
 one must study the transverse mass,
 \,$M_T^2(W_0Z_0)=[\sqrt{M^2(\ell\ell\ell)+p_T^2(\ell\ell\ell)}
    +|p_T^{\rm miss}|]^2-|p_T^{}(\ell\ell\ell)+p_T^{\;\rm miss}|^2$
 \cite{bagger}.
 We compute both the signal and backgrounds
 for the $M_T(W_0Z_0)$ distribution
 and depict them in Fig.\,\ref{sb-newcuts}.
 Counting the signal and background events
 in the range $0.85M_{W_1}<M_T<1.05M_{W_1}$,
 we further obtain the required integrated luminosities
 for $3\sigma$ and $5\sigma$ detections of the $W_1$ boson,
 as shown in Fig.\,\ref{IL}.

 \vspace*{3mm}
 \noindent
 {\bf 5.~Complementarity and the LHC Discovery Potential}
 \vspace{3mm}

 We have performed the first gauge-invariant study of
 LHC signatures of the new gauge bosons predicted by
 the Minimal Higgsless Model (MHLM)\,\cite{3site}.
 The $W_0Z_0Z_0$ channel is of special importance
 due to its distinct $jj4\ell$ signals and the full
 reconstruction of $W_1$ peak (Fig.\,1).
 We find that the simple cuts (\ref{supp-cuts})
 can sufficiently suppress all the SM backgrounds and
 single out the $jj4\ell$ signals.
 The $W_0Z_0jj$ channel has a larger cross section
 when $M_{W1}$ is heavy,
 but measuring the $W_1$ peak is harder
 due to the missing-$E_T$ of final state neutrinos.
 Hence, the $W_0Z_0Z_0$ channel plays a {\it crucial
 complementary role} to the $W_0Z_0jj$ channel for
 {\it co-discovering} $W_1$ bosons at the LHC.
 Confirming the $W_1$ signals in {\it both channels},
 as well as the {\it absence} of a Higgs-like signal in
 $pp\to Z_0Z_0qq\to 4\ell \,qq$,
 will be strong evidences for Higgsless EWSB.
 %

 \vspace*{3mm}
 We summarize the $3\sigma$ and $5\sigma$ detection potential of the
 LHC in Fig.\,\ref{IL}, where the required integrated luminosities
 are derived over the full range of allowed $W_1$ mass.
 Here we have included statistical error in determining the discovery
 potential; systematic error and other detector details are beyond the
 current scope.
 We find that, for $M_{W_1}=500\,(400)$\,GeV, the $5\sigma$
 discovery of $W_1$ requires an integrated luminosity
 of 26\,(7.8)\,fb$^{-1}$
 for $pp \to W_0Z_0Z_0\to jj\,4\ell$, and
 12\,(7)\,fb$^{-1}$ for
 $pp \to  W_0Z_0 jj \to \nu3\ell\,jj$.
 These discovery reaches will be achieved
 within the first few years' run at the LHC.

 \vspace*{6mm}
 \noindent
 {\bf 6.~Conclusions}
 \vspace{3mm}

In this work, we have presented the first realistic and consistent study of LHC signatures for the
Minimal Higgsless Model (MHLM), which is proven (through the unitary gauge and 't\,Hooft-Feynman gauge)
to be exactly gauge-invariant.
Such exact gauge-invariance was never demonstrated before in all other Higgsless studies.

In the first part of our work, we proposed and studied, for the first time, a new promising
channel $WZZ(\to jj4\ell)$ for detecting new $W_1$ bosons in the Higgsless model.
In the second part, we gave the first quantitative study of the $WZjj$ channel and its
LHC-signatures of $W_1$ from the exactly gauge-invariant MHLM Lagrangian.
In this channel we demonstrated for the first time the large cancellations between fusion and non-fusion
graphs in the MHLM by {\it consistently including all new physics contributions} [cf.\ Fig.\,2(a)-(b)].
We also proved that a non-gauge-invariant Higgsless toy model does violate this large cancellation, and
leads to erroneous high energy behavior of $M_{W_0Z_0}$ [cf. Fig.\,2(c)].
Therefore, it is vital to adopt exactly gauge-invariant Higgsless models such as the MHLM\,\cite{3site}
for correctly studying LHC-phenomenology via $2\to 4$ processes.
Furthermore, our study computed not only the realistic signal events
but also the correct background events (cf. Fig.\,3),
which are crucial for predicting the LHC discovery potential of new $W_1$ bosons.

We have further demonstrated 
the {\it complementarity between the $WZZ$ and $WZjj$
channels,} which are both very promising.
The LHC discovery potential of new $W_1$ bosons in both channels
was first quantitatively predicted over the full mass-range of $W_1$ in Fig.\,4.

 \vspace*{3mm}
 \noindent
 {\bf Acknowledgments:}~  \\ [1.5mm]
 We thank Bing Zhou for discussing the LHC/ATLAS detection,
 and Tao Han for discussing the QCD-scale used
 in Ref.\,\cite{HanWillenbrock}.
 This research was supported by the NSF of China under
 grants 1062552, 10635030, 90403017, 10705017; and by
 the US~NSF under grant PHY-0354226 as well as
 MSU High Performance Computing Center.
 The authors also acknowledge the support of
 the Radcliffe Institute of Advanced Study at Harvard University,
 the CERN Theory Institute and Fermilab Theory Department
 during the completion of this work.

 \vspace*{-3mm}

\end{document}